\documentclass[12pt]{article}
\usepackage{epsfig}
\begin{document}
\begin{center}
{\large\bf Symmetry and symmetry breaking in particle physics}\\
TSOU Sheung Tsun\\
Mathematical Institute, Oxford University\\
24--29 St.\ Giles', Oxford OX1 3LB\\
United Kingdom.\\
tsou\,@\,maths.ox.ac.uk
\end{center}

{\bf Abstract}\\

Symmetry, in particular gauge symmetry, is a fundamental principle in
theoretical physics.  It is intimately connected to the geometry of
fibre bundles.  A refinement to the gauge principle, known as
``spontaneous symmetry breaking'', leads to one of the most successful
theories in modern particle physics.  In this short talk, I shall try
to give a taste of this beautiful and exciting concept.

\vfill
\noindent{Invited talk at the 8th General Meeting, European Women in
Mathematics, 12--16 December, 1997, Trieste, Italy, to appear in the 
Proceedings.} 

\section{Introduction}
The concept of symmetry is one of the very few on which mathematicians
and physicists agree, namely that
\begin{center}
\fbox{{\footnotesize SYMMETRY}\ $\equiv$\ {\footnotesize GROUPS}}\,.
\end{center}
Hence we shall use these terms interchangeably.

In particle physics, there are two main uses of groups:
\begin{enumerate}
\item as transformation groups under which a theory is {\em
invariant};
\item as group representations for classifying the many particles we
see.
\end{enumerate}

In a sense, the first is all important, just like the main characters of a
play.  The second is more like the supporting cast, without which the
theory, although it can stand on its own, is much less interesting and
also much less realistic.

The next question is: which groups does one use or need?  Generally
speaking, finite-dimensional compact semi-simple Lie groups.  In this
talk, in order to simplify the presentation but without losing the
essentials, I shall consider almost exclusively only the following:
for abelian groups $U(1)$, and for nonabelian groups the unitary
groups $U(N)$ and $SU(N)$.  At the end I shall mention an example
where a discrete group figures.

\section{The particles: a lightning view}
Particles used to be called {\em elementary} particles, which made
good sense when we knew only the electron, the proton and the neutron,
and they were adequate for forming {\em all} the elements in the
Periodic Table.  Then Einstein proved the existence of the photon as a
particle.  Also Dirac postulated the existence of anti-particles,
which was well borne out by later experiments. \ldots All in all,
there are now more than 150 of them listed, and the number keeps on
increasing!  It would be highly unsatisfactory if we had to put them
all in one or more representations or `multiplets' without a good
theoretical guidance.

Fortunately, we do now have a theoretical basis, the gauge principle,
which we shall study in the next section.  In the light of the gauge
principle, particles can be classified under three headings:
\begin{itemize}
\item Vector bosons: $\gamma$ (the photon), $W^+,\ W^-,\ Z^0$.
\item Leptons: $e,\ \nu_e;\ \mu,\ \nu_\mu;\ \tau,\ \nu_\tau$.  (In
words, the electron, the electron neutrino, etc.)
\item Quarks: these are not observable themselves, but they form most
of the other particles by combining two or three together.  Each quark
$q$ is in
the 3-dimensional or fundamental representation, and directly
observable particles occur
in the 1-dimensional or singlet representation as follows:
$$\begin{array}{ccrr}
qqq: & {\bf 3} \otimes {\bf 3} \otimes {\bf 3} & = & {\bf 1}
\oplus \cdots \\
q \bar{q}:& {\bf 3} \otimes {\bar{\bf 3}} &=&{\bf 1}
\oplus \cdots
\end{array} $$
\end{itemize}

Note that only singlets can be observed as free particles, as will be
explained later.

\section{The gauge principle}
We said at the beginning that the invariance of a theory under certain
group transformations is the most important aspect of symmetry.  Let
us study it now in greater detail.

Recall classical electromagnetism.  The skew rank 2 field tensor
$F_{\mu\nu}\ (\mu,\nu=0,1,2,3)$ has as its components the electric
{\bf E} and magnetic {\bf B} fields:
$$ F_{\mu\nu} = \left(
\begin{array}{rrrr}
0&E_1 & E_2 & E_3 \\
-E_1 & 0 &-B_3 & B_2 \\
-E_2 & B_3 & 0 &-B_1 \\
-E_3 & -B_2 & B_1 & 0
\end{array} \right).  $$
These are directly measurable quantities and hence do not transform
under any symmetries.  However, one can and does introduce a vector
potential $A_\mu$, related to $F_{\mu\nu}$ by
$$ F_{\mu\nu} = \partial_\nu A_\mu - \partial_\mu A_\nu,  $$
so that there is a freedom in changing $A_\mu$ without affecting
$F_{\mu\nu}$:
$$ A_\mu \mapsto A_\mu + ie \partial_\mu \Lambda, $$
where $\Lambda (x)$ is a scalar field, and $e$ is a `coupling'
constant representing the strength of interaction.  In classical
theory, there is no need to consider the potential $A_\mu$.  However,
in quantum theory, it was demonstrated that $F_{\mu\nu}$ is not
enough to describe the physics and one {\em needs} $A_\mu$.  This is
the famous Bohm--Aharonov experiment.

The `gauge freedom' in $A_\mu$ is in fact linked to the arbitrary
phase of the electron wave function:
$$ \psi \mapsto  e^{ie\Lambda}\,\psi.  $$
Hence the relevant group for the symmetry of electromagnetism is:
\begin{center}
\fbox{$G=U(1)$}
\end{center}

In 1954, Yang and Mills extended this gauge principle to a nonabelian
group $G$:
\begin{eqnarray*}
A_\mu & \mapsto & S A_\mu S^{-1} - \frac{i}{g} (\partial_\mu S)
S^{-1},\\
\psi & \mapsto &S \psi,
\end{eqnarray*}
where $S \in G$.  

This is the famous Yang--Mills theory.  In the last 20 years or so, it
has been generally accepted that Yang--Mills theory is the basis of
{\em all} of particle physics:
\begin{center}
\fbox{{\footnotesize YANG--MILLS THEORY} $=$ {\footnotesize BASIS OF
ALL PARTICLE PHYSICS}}
\end{center}

A refinement of gauge symmetry is called {\em symmetry breaking},
where the whole theory (including equations of motion) is invariant
under a group $G$ but a particular solution (or `vacuum') is
invariant only under a subgroup $H \subset G$.  This will be important
for later applications.

\section{The geometry of gauge theory}
Although it was not realized at the time, gauge theory is intimately
linked with geometry.  In fact it is as geometric a theory as
Einstein's general relativity.  Table \ref{mathphys}, borrowed from a
paper by Yang, underlines this fact.

\begin{table}
\begin{center}
\begin{tabular}{|l|l|}  \hline  
{\em Physics} & {\em Mathematics}  \\ \hline
Special Relativity & Flat Space-time \\
General Relativity & Riemannian Geometry  \\
Quantum Mechanics & Hilbert Space \\
Electromagnetism and & Fibre Bundles \\
\ \ Yang-Mills Gauge Theory &  \\ \hline
\end{tabular}
\caption{Mathematics and physical theories}
\label{mathphys}
\end{center}
\end{table}
\vspace*{3mm}

\begin{figure} [h]
\centering
\includegraphics{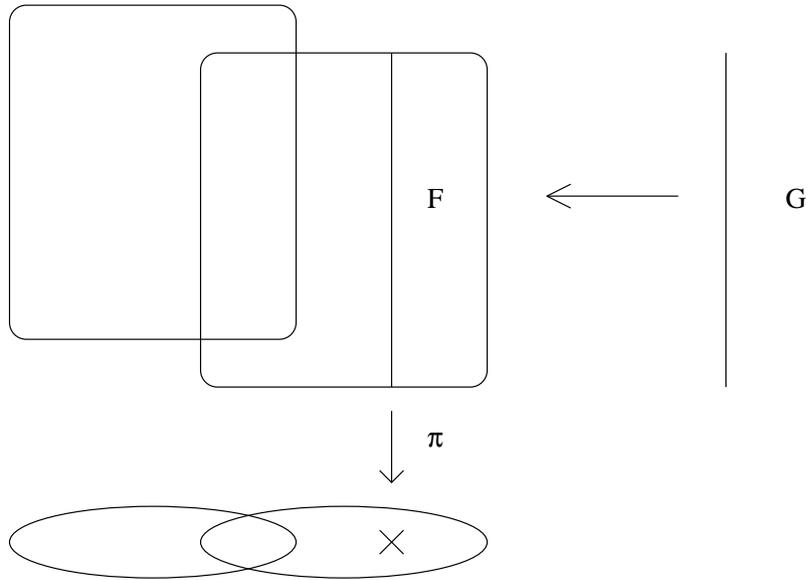}
\caption{Sketch of a principal bundle.}
\label{bundle}
\end{figure}

Recall the definition of a principal fibre bundle, as illustrated in
the accompanying sketch (Figure \ref{bundle}). 

Thus a principal fibre bundle consists of a manifold $P$ ({\em total
space}), a manifold $X$ ({\em base space} or {\em spacetime}), a 
projection $\pi$ and a
group $G$ ({\em structure} or {\em gauge group}).  Above any point $x
\in X$ the inverse image $\pi^{-1} (x) \subset P$ is called the
typical fibre $F$, and is homeomorphic to $G$.
Above an open set
$U_\alpha$ of $X$, the inverse image $\pi^{-1} (U_\alpha) \subset P$
is homeomorphic to the product $U_\alpha \times F$:
$$\phi_\alpha \colon U_\alpha \times F \to \pi^{-1} (U_\alpha).  $$
Thus in a sense, the manifold $P$ is a `twisted' product of $G$ and
$X$, the twisting being done by the action of the group:
\begin{eqnarray*}
\phi_{\alpha,x} & \colon & F \to \pi^{-1} (x) \\
&& y \mapsto \phi_\alpha (x,y),
\end{eqnarray*}
with
$$ \phi_{\beta,x}^{-1} \phi_{\alpha,x} \colon F \to F$$
giving the relevant action of the group $G$.

A trivial bundle is then just the product $X \times G$.  The most
well-known example of a nontrivial bundle is the M\"obius band, where
twisting is done by the 2-element group ${\bf Z}_2$.  An example which
is useful in physics is the {\em magnetic monopole}, which can be
represented topologically by $S^3$, which in turn is a nontrivial
$S^1$ bundle over $S^2$ (the Hopf bundle, of Chern class 1, for the 
experts).  Here
spacetime is thought of as $S^2 \times {\bf R}^2$, where the
second factor is just a vector space with no topology, and can thus be
ignored for the present purpose.  Ordinary electromagnetism without
magnetic monopoles is given topologically by the trivial bundle bundle
${\bf R}^4 \times S^1$.  In both cases, the typical fibre is the
circle $S^1$,
which is homeomorphic to the group $U(1)$.

To proceed further we need to introduce a {\em connection} on the
principal bundle $P$.  This is a 1-form $A$ on $P$ with values in the Lie
algebra {\bf g} of $G$, satisfying certain properties and giving a
prescription for differentiating vectors and tensors.  Locally it
combines with the usual partial derivative to give the {\em covariant
derivative}:
$$ D_\mu = \partial_\mu - i g [A_\mu, \cdot].  $$
In differential geometry and in gauge theory one has to replace the
partial derivative by the covariant derivative so as to preserve the
invariance or symmetry of the system.

From the connection one can define the {\em curvature}:
$$ F_{\mu\nu} = \partial_\nu A_\mu -\partial_\mu A_\nu +ig
[A_\mu,A_\nu]. $$
One recognizes immediately that these are respectively the gauge
potential and the gauge field introduced in the last section, where
the extra commutators (in the Lie algebra) take into account that
now the group is in general nonabelian.

With this language, the mechanism of symmetry breaking can be stated
as the case when the twisting of the bundle are by elements of a
subgroup $H$ of $G$, and when the connection 1-form takes values in
the corresponding Lie subalgebra.  One says then that the bundle with
connection is {\em reducible} to the subgroup $H$.  An important 
example is the 't Hooft--Polyakov magnetic monopole, which is a
nontrivial $U(1)$ reduction of a trivial $SU(2)$ bundle, given by the
exact sequence (for those who are fond of such things):
$$ \cdots \to \pi_2 (SU(2)) \to \pi_2 (SU(2)/U(1)) \to \pi_1 (U(1))
\to \pi_1 (SU(2)) \to \cdots.  $$
The first and last terms being zero, one gets the isomorphism
$$ \pi_2 (SU(2)/U(1)) \cong \pi_1 (U(1)). $$

\section{Briefest summary of the Standard Model}
Following the gauge principle, we can now try to fit the three types
of particles of section 2 into a more systematic pattern, the better
to exhibit their symmetry properties.

The vector bosons, also known as gauge bosons, are the potential
$A_\mu (x)$ when considered as fields.  Note that in the language of
quantum field theory, the concept of ``particles'' and ``fields'' are
interchangeable: particles interact by influencing the spacetime in
their neighbourhood and thus giving rise to fields, that is, functions
of spacetime with a definite tensor property (whether scalar, vector,
rank 2 skew tensor, etc.).  According to the interaction, we have a
specific symmetry or gauge group.  The other two types of particles
are usually thought of as ``matter fields'' belonging to
representations of the corresponding groups.

We now recognize that, other than gravitation, there are two
fundamental forces of Nature: the strong and the electroweak.  The
electroweak theory is an example of a gauge theory with symmetry
breaking.  The idea, called the Weinberg--Salam model, is that at high
energies when the Universe was much younger the symmetry was not
broken, but as the Universe cooled down the $U(2)$ gauge group broke
down to the $U(1)$ subgroup which is the electromagnetism of today.
The rest of the $U(2)$ interaction manifests itself in the present-day
weak interaction, of which radioactivity is the most commonly known
aspect.  The breaking also leaves some remnant fields called the Higgs
fields which are yet to be discovered.

As mentioned already, each quark is in a 3-dimensional representation
of $SU(3)$.  Hence a quark has in fact three states, fancifully called
{\em colour}.  This ``colour'' is not directly observable, as only
states in the singlet representation can exist free.  We say that the
$SU(3)$ symmetry is {\em exact} and {\em confined}.

Table \ref{particles} summarizes these ingredients of the so-called
Standard Model of particle physics.  The particles in square brackets are
not (or have not been) directly observed, but they are part of the theory.

\begin{table}
\begin{center}
\begin{tabular}{|l|c|c|c|} \hline
{\sc Force} & {\sc Group} & {\sc Gauge bosons} & {\sc Matter} \\
\hline \hline
Strong & $SU(3)$ & [Gluons] & [Quarks] \\ 
(QCD) &&&\\ \hline
Electroweak & $U(2)$ & $\gamma,W^{\pm},Z^0$ & Leptons \\
(Weinberg--Salam) &&& [Higgs] \\ \hline
\end{tabular}
\caption{Forces and Fields in the Standard Model}
\label{particles}
\end{center}
\end{table}

The standard model can in fact be schematically represented as:
\begin{center}
(QCD + Weinberg--Salam) $\times$ 3
\end{center}
the gauge group being $SU(3) \times SU(2) \times U(1)/{\bf Z}_6$.
Most physicists neglect the six-fold identification, but it is
important for identifying the correct particle representations.

The multiplication by 3 above is necessary to model another aspect of
the particle spectrum known as {\em generations}.  Take the charged
leptons as an example.  There are 3 of them: the electron $e$, the
muon $\mu$ and the tauon $\tau$.  Except for their very different
masses, they behave in extremely similar fashion.  The same pattern is
repeated for their neutral `partners' the neutrinos
$\nu_e,\nu_\mu,\nu_\tau$.   The quarks also come in three generations:
the `up' and `down' as the lightest generation, the `charm' and
`strange' as the next in mass, and the `top' and `bottom' as the
heaviest.  Table \ref{generations} arranges the 3 generations as 3
rows.  The subscripts L and R refer to the left-handed and
right-handed field components, a refinement we shall not have time to
go into.

\begin{table}
\begin{center}
\begin{tabular}{|ccl|rc|}
\multicolumn{3}{c}{\sc Quarks} & \multicolumn{2}{c}{\sc Leptons} \\
\hline
&&&&\\
$ \left( \begin{array}{c} 
u \\ d \end{array} \right)_L$ & $u_R$ & $d_R$ \quad & \quad 
$ \left( \begin{array}{c} 
\nu_e \\ e \end{array} \right)_L$ & $e_R$ \\ 
&&&& \\ \hline
&&&&\\
$ \left( \begin{array}{c} 
c \\ s \end{array} \right)_L$ & $c_R$ & $s_R$ \quad & \quad 
$\left( \begin{array}{c} 
\nu_\mu \\ \mu  \end{array} \right)_L$ & $\mu_R$ \\ 
&&&&\\ \hline
&&&&\\
$\left( \begin{array}{c}
t \\ b \end{array} \right)_L$ & $t_R$ & $b_R$ \quad & \quad 
$ \left( \begin{array}{c}
\nu_\tau \\ \tau  \end{array} \right)_L$ & $\tau_R$ \\ 
&&&&\\ \hline
\end{tabular}
\caption{Generations of Quarks and Leptons}
\label{generations}
\end{center}
\end{table}

The role of the Higgs fields in the standard model is crucial.  They
break the $U(2)$ symmetry, give masses to the gauge bosons $W,Z$ and
also give masses to the quarks and charged leptons.  Without them, all
particles would be massless.  Notice that the neutrinos are supposed
to be massless, although some recent experiments in particle physics
and astrophysics indicate that they may have extremely small masses.

Even with this briefest of summaries of the Standard Model we can
already see how symmetry plays a crucial organizing role in our
understanding of particle physics.  And in this {\em gauge} symmetry
is of prime importance.

\section{Electric--magnetic duality: example of a discrete symmetry}
It is well-known that electromagnetism has a discrete ${\bf Z}_2$
symmetry, that is, the equations are invariant under the change from
`electric' to `magnetic' and vice versa.  Let us look at this in a
little more detail.

As described in section 3, we can start with the potential $A_\mu$ and
define the field tensor $F_{\mu\nu}$ by
$$F_{\mu\nu} = \partial_\nu A_\mu - \partial_\mu A_\nu.$$
Further introduce the Hodge star operator, which in this case goes
from 2-forms to 2-forms:
$${}^*\!F_{\mu\nu}=-\frac{1}{2} \epsilon_{\mu\nu\rho\sigma}
F^{\rho\sigma}. $$
This operation interchanges electric fields and magnetic fields.  We
then have the identity:
$$ \partial_\mu {}^*\!F^{\mu\nu} =0, $$
which always holds for $F_{\mu\nu}$ defined as above in terms of an
$A_\mu$.  On the other hand, by Gauss' theorem, this `divergen--free'
condition is equivalent to the absence of magnetic monopoles, because
${}^*\!F_{\mu\nu}$ gives the magnetic flux out of such an object if
present.  This very significant link between a geometric statement and
a physical statement can be schematically represented as:
$$\underbrace{A_\mu {\rm \ exists}}_{\rm geometry}\ \  \stackrel{\rm
Poincar\acute{e}}{\Longleftrightarrow}\ \  \partial_\mu \mbox{}^*\! F^{\mu\nu}
=0\ \  \stackrel{\rm Gauss}{\Longleftrightarrow} \ \ \underbrace{\rm no\
magnetic\ monopoles}_{\rm physics} $$
In the language of differential forms, the geometric statement is
no other than
$$ F\ {\rm exact}\ \  \stackrel{\rm locally}{\Longleftrightarrow}\ \
F\ {\rm closed} $$

Now in the absence of electric charges (remember: only the main
characters and no supporting cast!), we have
$$ \partial_\mu F^{\mu\nu} = 0,$$
just as for the case of magnetic monopoles above, only this time we
have $F^{\mu\nu}$ instead of ${}^*\!F^{\mu\nu}$.  So we have the
`dual' of the scheme above:
$$\underbrace{\tilde{A}_\mu {\rm \ exists}}_{\rm geometry}\ \  \stackrel{\rm
Poincar\acute{e}}{\Longleftrightarrow}\ \  \partial_\mu F^{\mu\nu}
=0\ \  \stackrel{\rm Gauss}{\Longleftrightarrow} \ \ \underbrace{\rm no\
electric\ charges}_{\rm physics} $$
We see that the electric--magnetic discrete symmetry indeed holds.

It can further be shown that electromagnetism is dual symmetric in the
above sense even in the presence of charges.

What is even more interesting---and this is what I am currently
working on---is that Yang--Mills theory (or
nonabelian gauge theory) is also dual symmetric, but the proof is not
all that straightforward.  One has to use techniques involving
infinite-dimensional loop variables and the dual transform is no
longer just the Hodge star but a loop space generalization of it.
What is interesting, and intriguing, is that this {\em discrete}
symmetry is clearly linked to the {\em continuous} gauge symmetry.
One consequence is that the gauge symmetry is now doubled: 
$$G \times \tilde{G},$$
where as groups the two factors are identical, only the physical
aspects they refer to are not identical but dual to each other.  Now
't Hooft proved a theorem which can be stated as follows: the $G$
symmetry is exact and confined if and only if the $\tilde{G}$ symmetry
is broken and massive.   Compare this to the actual symmetries of the
Standard Model:
\begin{center}
\begin{tabular}{rl}
$SU(3)$ & exact and confined \\
$U(2)$ & broken and massive
\end{tabular}
\end{center}
Applying 't Hooft's theorem to these symmetries lead to very
interesting consequences which I do not have time to talk about.

\section{Conclusions}
Let me summarize the salient points about symmetry in particle physics
that I have mentioned:
\begin{enumerate}
\item Symmetry is all important in physics.  For lack of time (and
expertise) I have omitted to treat many symmetries, such as Lorentz
symmetry, diffeomorphism symmetry, supersymmetry, \ldots.
\item There are two main uses of groups:
\begin{enumerate}
\item in the gauge principle as invariance, and
\item for particle classification using representations.
\end{enumerate}
\item The Standard Model is a triumph of the gauge principle.
\item Electric--magnetic duality (a discrete ${\bf Z}_2$ symmetry),
when generalized to Yang--Mills theory, leads to very interesting
results.
\end{enumerate}

If, however, you wish  to take away with you just one point, then
I recommend:
\begin{center}
\fbox{{\footnotesize SYMMETRY}\ $\equiv$\ {\footnotesize GROUPS}}\,.
\end{center}

\section*{Acknowledgements}
I thank Bodil Branner and Sylvie Paycha for inviting me to this
meeting, and the British Branch of EWM for a travel grant.

\section*{References}
There are many excellent textbooks and semi-popular books on modern
particle physics which emphasize its symmetry properties. 
There are also excellent articles in {\em Scientific
American} which are most suitable to give a taste of the beauty of the
subject.  Below is a random selection of such, the first being a more
general appreciation of symmetry in physics by the originator of
Yang--Mills theory:
\begin{enumerate}
\item Chen Ning Yang, Symmetry and Physics, in {\em Oskar Klein Memorial
Lectures}, Vol.\ 1, World Scientific Publishing Company, Singapore, 1991.
\item G.\ 't Hooft, {\em Scientific American}, June 1980, p.\ 90.
\item S.\ Weinberg, {\em Scientific American}, June 1981, p.\ 52; {\em
ibid.} July 1974, p.\ 50.
\item C.\ Quigg, {\em Scientific American}, April 1985, p.\ 64.
\item S.\ Weinberg, {\em Dreams of a Final Theory}, Hutchison 1993.
\end{enumerate}

For the reader who might want to know more about the last part of this
lecture, here are a few of my recent articles (the last with an amusing
application from the Serret--Frenet formulae for space curves):
\begin{enumerate}
\item Chan Hong-Mo and Tsou Sheung Tsun, Physical Consequences of 
  Nonabelian Duality in the Standard
  Model, hep-th/9701120, {\em Phys.\ Rev.} {\bf
  D57}\,(1998)\,2507--2522. 
\item Chan Hong-Mo and Tsou Sheung Tsun, Standard Model with Duality: 
  Theoretical Basis, 
  hep-th/9712171;  Standard Model with Duality: Physical Consequences, 
  hep-ph/9712436;
  invited lectures at the Cracow Summer
  School on Theoretical Physics, May--June 1997, Zakopane,
  {\em Acta Phys.\ Pol.} {\bf B28}\,(1997)\,3027--3040; 3041--3056.
\item Jos\'e Bordes, Chan Hong-Mo, Jakov Pfaudler and Tsou Sheung Tsun,
  Features of Quark and Lepton Mixing from Differential Geometry
  of Curves on Surfaces, hep-ph/9802436, February 1998.
\end{enumerate}
\end{document}